\documentclass[]{aastex631}

\usepackage{rotating}
\newcommand{\ie}{{\it i.e.}}
\newcommand{\eg}{{\it e.g.}}

\newcommand{\kepler}{{Kepler}}
\newcommand{\Kepler}{{Kepler}}

\newcommand{\gaia}{{\it Gaia}}

\newcommand{\vx}{$v_{\bf x}$}
\newcommand{\vy}{$v_{\bf y}$}
\newcommand{\vz}{$v_{\bf z}$}
\newcommand{\vb}{$v_{\bf b}$}
\newcommand{\x}{${\bf x}$}
\newcommand{\y}{${\bf y}$}
\newcommand{\z}{${\bf z}$}
\newcommand{\kms}{km s$^{-1}$}

\newcommand{\mura}{$\mu_\alpha$}
\newcommand{\pmra}{$\mu_\alpha$}
\newcommand{\mudec}{$\mu_\delta$}
\newcommand{\pmdec}{$\mu_\delta$}
\newcommand{\parallax}{$\pi$}
\newcommand{\ra}{$\alpha$}
\newcommand{\dec}{$\delta$}

\newcommand{\python}{{\it Python}}

\newcommand{\vxprecision}{4}
\newcommand{\vyprecision}{18}
\newcommand{\vzprecision}{4}
\newcommand{\dprecision}{1}
\newcommand{\nstars}{148,590}
\newcommand{\ngaia}{23,013}
\newcommand{\nlamost}{22,420}
\newcommand{\napogee}{7,697}
\newcommand{\nrv}{38,884}
\newcommand{\kicstar}{12218729}
\newcommand{\nfun}{198,451}

\begin{document}

\title{The 3D Galactocentric velocities of Kepler stars: marginalizing
over missing RVs}

% \correspondingauthor{Ruth Angus}
% \email{rangus@amnh.org}

\author[0000-0003-4540-5661]{Ruth Angus}
\affiliation{Department of Astrophysics, American Museum of Natural History,
200 Central Park West, Manhattan, NY, USA}
\affiliation{Center for Computational Astrophysics, Flatiron Institute,
162 5th Avenue, Manhattan, NY, USA}
\affiliation{Department of Astronomy, Pupin Hall, Columbia University,
Manhattan, NY, USA}

\author[0000-0003-0872-7098]{Adrian M. Price-Whelan}
\affiliation{Center for Computational Astrophysics, Flatiron Institute,
162 5th Avenue, Manhattan, NY, USA}

\author[0000-0002-7550-7151]{Joel C. Zinn}
\altaffiliation{NSF Astronomy and Astrophysics Postdoctoral Fellow.}
\affiliation{Department of Astrophysics, American Museum of Natural History,
Central Park West at 79th Street, New York, NY 10024, USA}

\author[0000-0001-9907-7742]{Megan Bedell}
\affiliation{Center for Computational Astrophysics, Flatiron Institute,
162 5th Avenue, Manhattan, NY, USA}

\author[0000-0003-4769-3273]{Yuxi (Lucy) Lu}
\affiliation{Department of Astronomy, Pupin Hall, Columbia University,
Manhattan, NY, USA}
\affiliation{Department of Astrophysics, American Museum of Natural History,
200 Central Park West, Manhattan, NY, USA}

\author[0000-0002-9328-5652]{Daniel Foreman-Mackey}
\affiliation{Center for Computational Astrophysics, Flatiron Institute,
162 5th Avenue, Manhattan, NY, USA}

\begin{abstract}
Precise Gaia measurements of positions, parallaxes, and proper motions provide
    an opportunity to calculate 3D positions and 2D velocities (\ie\
    5D phase-space) of Milky Way stars.
Where available, spectroscopic radial velocity (RV) measurements provide full
    6D phase-space information, however there are now and will remain many
    stars without RV measurements.
% Gaia will provide RVs for stars as faint as the 15th magnitude in its third
%     data release, however there are now and will remain many stars without RV
%     measurements.
Without an RV it is not possible to directly calculate 3D stellar velocities,
    however one can {\it infer} 3D stellar velocities by marginalizing over
    the missing RV dimension.
In this paper, we infer the 3D velocities of stars in the
    Kepler field in Cartesian Galactocentric coordinates (\vx, \vy, \vz).
    We directly calculate velocities for around a quarter of all Kepler
    targets, using RV measurements available from the Gaia, LAMOST and APOGEE
    spectroscopic surveys.
Using the velocity distributions of these stars as our prior, we infer
    velocities for the remaining three-quarters of the sample by marginalizing
    over the RV dimension.
% For many applications, including kinematic age-dating, precise velocities in
%     the \vx\ and \vz\ directions are sufficiently useful.
% In this paper, we infer \vx\ and \vz\ with typical uncertainties of
%     \vxprecision\ and \vzprecision\ \kms\ for a sample of \nstars\ in the
%     Kepler field, by marginalizing over unknown RV.
% The Kepler field lies at a low Galactic latitude and is closely aligned with
%     the y-axis of the Cartesian Galactocentric coordinate system.
% This means that, without an RV, \vy\ is poorly constrained but \vx\ and \vz\
%     can be precisely inferred.
The median uncertainties on our inferred \vx, \vy, and \vz\ velocities are
    around \vxprecision, \vyprecision, and \vzprecision\ kms$^{-1}$,
    respectively.
We provide 3D velocities for a total of \nstars\ stars in the Kepler field.
These 3D velocities could enable kinematic age-dating, Milky Way stellar
    population studies, and other scientific studies using the benchmark
    sample of well-studied Kepler stars.
Although the methodology used here is broadly applicable to targets across the
    sky, our prior is specifically constructed from and for the Kepler field.
Care should be taken to use a suitable prior when extending this method to
    other parts of the Galaxy.
\end{abstract}


\section{Introduction}

Gaia has revolutionized the field of Galactic dynamics by providing positions,
parallaxes and proper motions with unparalleled precision for a large number
of Milky Way stars.
So far, Gaia has provided positions, parallaxes and proper motions for around
1.7 billion stars, and radial velocities (RVs) for more than 7 million stars
across its 1st, 2nd and early-3rd data releases \citep{gaia, gaia_dr2,
gaia_edr3}.
In combination, proper motion, position, and RV measurements provide full 6D
phase-space information for any given star, which can be used to calculate its
Galactic orbit.
% 3D stellar velocities are useful for calculating the relative space-motions of
% stars, and for revealing their Galactic orbits.
The orbits of stars are useful for kinematic age-dating, for exploring the
secular dynamical evolution of the Galaxy, for differentiating between nascent
and accreted stellar populations in the Milky Way's halo, and many other
applications.

% Motivation
One particular motivation
% for calculating the velocities of Kepler targets
is to use Galactic kinematics to study stellar evolution, either by using
vertical velocity dispersion as an age proxy, or by calculating stellar ages
via an age-velocity dispersion relation \citep[\eg][]{angus2020, lu2021}.
The ages of stars, particularly GKM stars on the main sequence, are difficult
to measure because their luminosities and temperatures evolve slowly
\citep[see][for a review of stellar ages]{soderblom2010}.
Galactic kinematics provides an alternative, statistical dating method.

% The star forming molecular gas clouds observed in the Milky Way have a low
% out-of-plane, or vertical, velocity \citep[\eg][]{stark1989, stark2005,
% aumer2009, martig2014, aumer2016}.
% In contrast, the vertical velocities of older stars are observed to be larger
% in magnitude on average \citep{stromberg1946, wielen1977, nordstrom2004,
% holmberg2007, holmberg2009, aumer2009, casagrande2011, yu2018, ting2019}.
% Galaxy formation simulations indicate that stars initially form from
% dynamically hot gas, which settles into a cooler thin disk over time during
% the early stages of Galactic evolution \citep[\eg][]{bird2013}.
% After this initial cooling phase, giant molecular clouds and Galactic spiral
% arms dynamically `reheat' the orbits of stars over time.
% Regardless of the exact history of dynamical cooling and heating,.
Older populations of stars are observed to have larger velocity dispersions
than younger populations, and this is generally thought to be caused by
dynamical heating of the Galactic disc by giant molecular clouds and spiral
arms \citep[\eg][]{stromberg1946, wielen1977, nordstrom2004, holmberg2007,
holmberg2009, aumer2009, casagrande2011, yu2018, ting2019}.
This behavior is codified by empirically-calibrated Age-Velocity dispersion
Relations (AVRs), which typically express the relationship between age and
velocity dispersion as a power law: $\sigma_v \propto t^\beta$, with free
parameter, $\beta$ \citep[\eg][]{holmberg2009, yu2018, mackereth2019}.
These expressions can be used to calculate the ages of stellar populations
from their velocity dispersions.
However, AVRs are usually calibrated in 3D Galactocentric velocities, and most
commonly in vertical velocity: \vz\ or W.
Regardless of the coordinate system, some transformation from RV and proper
motion in equatorial coordinates is usually required to calculate the
kinematic ages of stars using an AVR.

RV measurements, combined with positions, parallaxes, and proper motions
measured in the plane of the sky, complete the full set of information needed
to calculate 3D stellar velocities.
However, RV generally has to be measured from a stellar spectrum -- an
observation that requires a significant number of photons and is thus
expensive to obtain, particularly for faint stars \footnote{RVs can also be
derived from perspective acceleration for high proper motion stars
\citep[\eg][]{lindegren2021}.}.
Fortunately however, the bulk circular velocity of the Galactic disc allows
stellar velocities to be inferred by using an informative prior that is
constructed from the velocities of stars with full 6D phase-space information.
% motions of stars in the Milky Way can Gaia proper motion measurements are of such incredible
% precision that, even without an RV measurement, the 3D velocity of a star can
% still be inferred by marginalizing over radial velocity.
This will often result in a velocity that is not equally well-constrained in
every direction, \ie\ the probability density function of a star's velocity
will be an oblate spheroid in 3D.
In the equatorial coordinate system, a star's velocity will be tightly
constrained in the directions of RA and dec, and only constrained by
the prior in the radial direction.
Transforming to any other coordinate system, a star's velocity probability
density function will change shape via a transformation that depends on its
position.

% , however for kinematic age-dating vertical velocity, \vz, is most
% useful.
% % For example, as described above the {\it vertical} velocities of stars (\vz\
% % or $W$) are often used to study the secular orbital heating of stars in the
% % Milky Way's disk \citep[\eg][]{beane2018, yu2018, ting2019, mackereth2019}.
% Unless the radial velocity direction precisely coincides with the vertical
% axis of the Milky Way, \ie\ a star lies along the $Z$ axis of the
% Galactocentric coordinate system, we can still extract some \vz\ information
% from Gaia proper motions alone.
% Of course, the lower the Galactic latitude of a star, the better the
% constraint on its \vz\ will be (hence why the Kepler field, located at low
% latitude, is particularly useful for vertical velocity studies).
% In some cases, the velocities of stars in particular directions are more
% useful than others, for example, the {\it vertical} velocities of stars (\vz\
% or $W$) are often used to study the secular orbital heating of stars in the
% Milky Way's disk \citep[\eg][]{beane2018, yu2018, ting2019, mackereth2019}.
% Unless the radial velocity direction precisely coincides with the vertical
% axis of the Milky Way, \ie\ a star lies along the $Z$ axis of the
% Galactocentric coordinate system, we can still extract some \vz\ information
% from Gaia proper motions alone.
% Of course, the lower the Galactic latitude of a star, the better the
% constraint on its \vz\ will be (hence why the Kepler field, located at low
% latitude, is particularly useful for vertical velocity studies).

In this work, we provide 3D velocities in \vx, \vy, and \vz\ for Kepler
targets.
Our motivation is chiefly to calculate vertical velocities (\vz) which can
then be used to calculate the ages of stellar populations via an AVR, from
which other empirical age-dating methods can be calibrated.
For example, empirical or semi-empirical models that relate the magnetic
activity or rotation periods of stars to their age can be used to infer the
ages of some low-mass dwarfs \citep[\eg][]{skumanich1972, barnes2003,
barnes2007, mamajek2008, matt2012, angus2019, claytor2020}, however, these
empirical relations are often poorly calibrated for low-mass and old stars
\citep[\eg][]{angus2015, vansaders2016, vansaders2018, metcalfe2019,
curtis2020, spada2019, angus2020}.
In \citet{angus2020} we used the velocities of Kepler stars in the direction
of Galactic latitude, \vb\, as a proxy for vertical velocity.
\vb\ can be calculated without an RV and it is similar to \vz\ for many Kepler
stars because the Kepler field lies at low Galactic latitude.
We used the \vb\ velocity dispersions of stars as an age proxy to explore the
evolution of stellar rotation rates.
In \citet{lu2021} we used {\it vertical} velocity dispersion (\vz) to
calculate kinematic ages for Kepler stars with measured rotation periods using
an age-velocity dispersion relation (AVR).
Those vertical velocities were inferred by marginalizing over missing RVs
using the method we describe in this paper.
To expand upon that work and provide an opportunity to apply kinematic
age-dating to more stars, we here calculate the 3D velocities of {\it all}
Kepler targets.
Although we focus on the Kepler field, the methodology presented in this paper
is applicable to stars across the sky if a suitable prior is used.

There are several other applications for which the 3D velocity of a star is
useful, even if its velocity is not equally well-constrained in every
direction.
For example, \citet{oh2017} used Gaia proper motions to identify comoving
pairs and groups of stars by marginalizing over missing RVs.
In their study, the relative space motions of pairs of stars were used to
establish whether they qualified as `comoving'.
In a pathological case where two stars (nearby on the sky) have near identical
proper motions and completely different RVs, their method would incorrectly
flag them as comoving stars, however in general the Gaia proper motion
precision is sufficiently high to make these cases rare.

Another work that predicts 3D velocities of Gaia targets is
\citet{dropulic2021}, in which the velocities of mock Gaia stars are predicted
with a neural network.
The network is trained on the velocities of stars {\it with} RVs and used to
predict the velocities of stars without.
The method we present here seeks to solve the same problem via a different
methodology: we use Bayesian inference instead of a neural network.
It is difficult to draw a direct comparison between their method and ours
because we use different coordinate systems (we use Cartesian, and they use
cylindrical coordinates), and because we focus on different populations --
they predict velocities for the entire Gaia catalog, whereas we concentrate on
a small part of the sky.
The two approaches will be useful for different scientific applications, and
it is extremely useful to have multiple approaches to solving this fundamental
problem in astronomy.

It is worth noting that ignoring the RV dimension and attempting to
calculate stellar velocities with proper motions and distances alone, \ie\
assuming that RV can be set to zero, will introduce a significant bias in the
resulting velocities.
We explore this point further in section \ref{sec:results}.

This paper is laid out as follows.
In section \ref{sec:data} we describe the data used in this paper.
In section \ref{sec:method} we describe how we calculate the kinematic ages of
Kepler stars from their positions, proper motions and parallaxes,
marginalizing over missing RVs.
We also justify the choice of prior probability density function (PDF).
In section \ref{sec:results} we present the 3D velocities of a subsample  of
the total \nstars\ Kepler stars and explore the accuracy and precision of our
method.

\section{The Data}
\label{sec:data}

We used the Kepler-Gaia cross-matched catalog that is available at
https://gaia-kepler.fun.
This catalog contains \nfun\ Kepler targets, cross-matched with Gaia targets
within in a 1'' radius and includes positions, parallaxes, and proper motions
from Gaia EDR3 and RVs from Gaia DR2.
We crossmatched this catalog with the catalog of photogeometric
distances inferred from Gaia EDR3 parallaxes \citep{bailer-jones2021} and
applied corrections to EDR3 parallaxes based on \citet{lindegren2021b}.
% created using the cross-match service and Vizier catalogue access tool
% provided by CDS, Strasbourg, France, as well as the astroquery and astropy
% python packages.
We also crossmatched this catalog with the LAMOST DR5 catalog and the APOGEE
DR16 stellar catalog \citep{cui2012, apogee_dr16, xiang2019}.
We removed stars with angular separations larger than 150 milliarcseconds
during each of these crossmatches.
Stars with effective temperatures that differed by more than 500K between
Gaia, APOGEE, and LAMOST were removed from the sample to minimize incorrect
crossmatches.
To remove stars with multiple crossmatches within 150 milliarcseconds, we only
kept the star with the smallest angular separation.
We also removed stars with a Gaia parallax $<$ 0, parallax signal-to-noise
ratio $<$ 10, and Gaia astrometric excess noise $>$ 5.
To preferentially select single stars, we removed stars with {\tt ruwe} $\geq$
1.4, {\tt ipd\_frac\_multi\_peak} $>2$, and {\tt
ipd\_gof\_harmonic\_amplitude} $\geq$ 0.1.
APOGEE reports VSCATTER, which is the error-deconvolved scatter in the
individual time-series RV measurements for each source and can indicate that a
star is a binary.
We removed 208 stars in our final sample with APOGEE VSCATTER greater than 1
\kms.
After applying these cuts our total number of targets was \nstars.
In total, \nrv\ stars in our sample have at least one RV measurement from
Gaia, LAMOST, or APOGEE; \ngaia\ have RVs from Gaia DR2, \nlamost\ from LAMOST
DR5, and \napogee\ from APOGEE DR16.
The APOGEE survey \citep[R $=$ 22,500;][]{apogee} has a higher spectral
resolution than Gaia \citep[R $=$ 11,500;][]{cropper2018}, which in turn is
higher than LAMOST \citep[R $=$ 1,800;][]{zhao2012}.
The median RV uncertainty for stars in our sample is around 0.1 km/s for
APOGEE RVs, 1 km/s for Gaia RVS, and 5 km/s for
LAMOST RVs.
In cases where stars had two or more available RV measurements, we adopted
APOGEE RVs as a first priority, followed by Gaia, then LAMOST.

\begin{figure}[ht!]
\caption{
    % The apparent magnitude (left) and temperature (right) distributions of
    % stars in our sample, with and without RV measurements from \gaia\ and
    % \lamost.
    The distribution of apparent Gaia magnitudes for
    stars in our sample with and without RV measurements from Gaia, LAMOST and
    APOGEE.
}
  \centering \includegraphics[width=.5\textwidth]{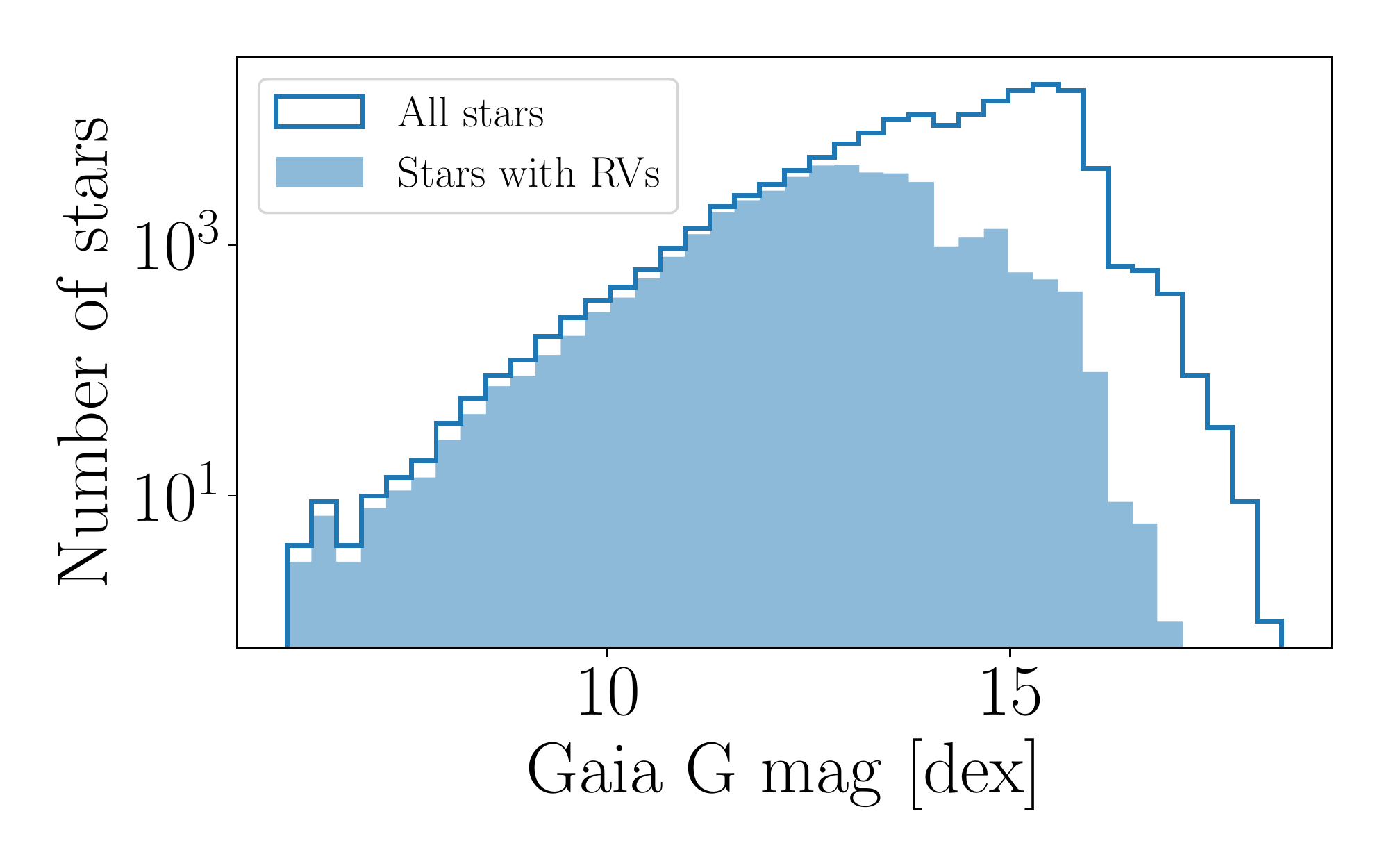}
\label{fig:rv_histogram}
\end{figure}

Although RVs are available for more than one in three Kepler targets, most
stars with RV measurements are bright.
Very few of the faintest stars have RVs because of the selection functions of
spectroscopic surveys.
% In our sample, one in 2.5 stars hotter than 5000 K had RV measurements,
% whereas only one in six stars cooler than 5000 K had RVs.
Most of the stars in our sample with Gaia RV measurements are brighter than
around 14th magnitude in Gaia $G$-band, and stars with LAMOST or APOGEE RVs
are mostly brighter than around 16th magnitude.
Figure \ref{fig:rv_histogram} shows the apparent magnitude distributions of
the stars in our sample, with and without RVs.
This figure reveals the combined selection functions of the Gaia, LAMOST and
APOGEE RV surveys and shows that faint stars are less likely to have RV
measurements than bright ones.

\begin{figure}[ht!]
\caption{
    A magnitude-temperature diagram of stars in the Kepler field with (left)
    and without (right) RVs provided by Gaia, LAMOST and APOGEE.
    The stars with RVs are generally hotter and more luminous than those
    without RVs, and include a large number of red clump stars and red giant
    branch stars.
    Stars without RVs are mostly concentrated on the main sequence.
}
  \centering \includegraphics[width=1\textwidth]{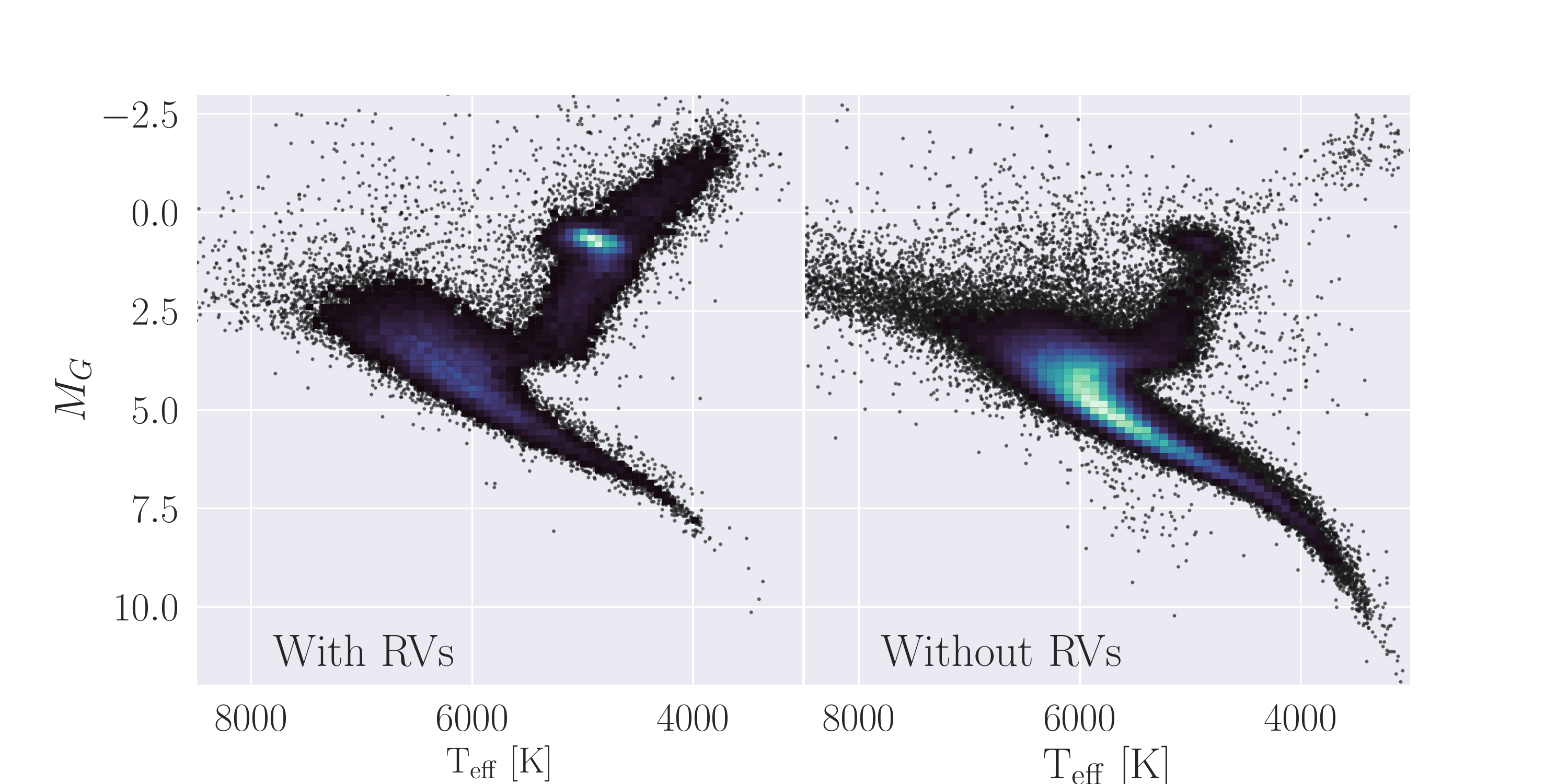}
\label{fig:CMD}
\end{figure}

To illustrate how the populations of stars with and without RVs differ, we
plot them on a color-magnitude diagram (CMD) in figure \ref{fig:CMD}.
The stars with RVs are generally hotter and more luminous than stars without.
Most stars with RVs fall on the upper main sequence, the red giant branch, and
the red clump.
Most stars without RVs fall on the main sequence.
This overall selection function is a combination of the APOGEE, LAMOST and
Gaia DR2 selection functions.

In this paper, we construct a prior using stars with RV measurements which we
then use to infer the velocities of stars without RV measurements.
However, given that the populations of stars with and without RVs are so
different, this could bias the velocities we infer, particularly if they are
prior-dependent.
We investigate this idea in section \ref{sec:prior} and find that the \vx\ and
\vz\ velocities we infer are relatively insensitive to the prior and therefore
unlikely to biased, however the \vy\ velocities we infer should be used with
caution as they are relatively prior-dependent.

\section{Method}
\label{sec:method}

In this section we describe how we calculate full 3D velocities for stars in
the Kepler field.
Around 1 in 3 Kepler targets have an RV from either Gaia, LAMOST, or APOGEE.
For these \nrv\ stars we calculated 3D velocities using the {\tt coordinates}
library of {\tt astropy} \citep{astropy2013, astropy2018}.
This library performs a series of matrix rotations and translations to convert
stellar positions and velocities in equatorial
% , or International Celestial Reference System (ICRS)
coordinates into positions and velocities in Galactocentric coordinates.
It converts positions, proper motions, parallaxes/distances, and RVs into \x,
\y, \z, \vx, \vy, \vz.
We adopted a Solar position of $r_\odot = 8.122$ kpc \citep{gravity2018} and
$z_\odot = 20.8$ pc \citep{bennet2019}, and a Solar velocity of $v\odot =
(12.9, 245.6, 7.78)$ \kms \citep{drimmel2018}.
For stars {\it without} RVs, we inferred their velocities by marginalizing
over their RVs using the method described below.

% It has been demonstrated that the dispersion in vertical velocity, \vz, of a
% population of stars increases with the age of that population,
% \citep[\eg][]{stromberg1946, wielen1977, nordstrom2004, holmberg2007,
% holmberg2009, aumer2009, casagrande2011, ting2019, yu2018}.
% Most AVRs are calibrated using velocities in Galactocentric coordinates, \vx,
% \vy\ and \vz, which can only be calculated with full 6-D position and velocity
% information, \ie\ proper motions, position and radial velocity.
% In \citet{angus2020} we explored rotational evolution using velocity
% dispersion as an age proxy, however we used velocity in the direction of
% Galactic latitude, \vb, instead of \vz.
% This is because \vb\ can be calculated without an RV measurement but is a
% close approximation to \vz\ for \kepler\ stars due to the orientation of the
% Kepler field.
% The \kepler\ field lies at low Galactic latitudes, ($\sim 5-20$\degrees), so
% the ${\bf z}$-direction is similar to the ${\bf b}$-direction for \kepler\
% stars.
% However, even at such low latitudes, kinematic ages calculated with \vb\
% instead of \vz\ are likely to be systematically larger because of mixing
% between \vz, \vx\ and \vy.
% A direct measurement or precise estimate of \vz\ is necessary to calculate
% accurate kinematic ages.

\subsection{Inferring 3D velocities (marginalizing over missing RV
measurements)}
\label{sec:inference}

% Three-dimensional velocities in galactocentric coordinates: \vx, \vy, and \vz\
% can only be directly computed via a transformation from 3D velocities in
% another coordinate system, like the equatorial coordinates provided by \gaia:
% \mura, \mudec, and RV.
% For stars with no measured RV in \gaia\ DR2, \vx, vy, and \vz\ can still be
% inferred from positions and proper motions alone, by marginalizing over
% missing RV measurements.
For each star in our sample without an RV measurement, we inferred \vx, \vy,
and \vz\ from the 3D positions -- RA (\ra), dec (\dec), and parallax
(\parallax), and 2D proper motions (\mura\ and \mudec) provided in the \gaia\
EDR3 catalog \citep{gaia_edr3}.
We also simultaneously inferred distance (instead of using inverse-parallax)
to model velocities \citep[see \eg][]{bailer-jones2015, bailer-jones2018}.

Using Bayes rule, the posterior probability of the velocity parameters given
the Gaia data can be written:
\begin{equation}
    p({\bf v_{xyz}}, D | \mu_{\alpha}, \mu_{\delta}, \alpha, \delta, \pi) =
    p(\mu_{\alpha}, \mu_{\delta}, \alpha, \delta, \pi | {\bf v_{xyz}}, D)
    p({\bf v_{xyz}}) p(D),
\end{equation}
where $D$ is distance and ${\bf v_{xyz}}$ is the 3D vector of velocities.
To evaluate the likelihood function, our model predicts observable data from
model parameters, \ie\ it converts \vx, \vy\, \vz\ and $D$ to \pmra, \pmdec\
and \parallax.
In the first step of the model evaluation, cartesian coordinates, \x, \y, and
\z\, are calculated from \ra, \dec, and $D$ by applying a series of matrix
rotations, and a translation to account for the Solar position.
The cartesian Galactocentric velocity parameters, \vx, \vy, and \vz, are then
converted to equatorial coordinates, \pmra\ and \pmdec\ via another rotation.
The posterior PDFs of the parameters \vx, \vy, \vz, and $\ln(D)$ are sampled
by evaluating this model over a range of parameter values which are chosen by
via the No U-Turns Sampler (NUTS) algorithm in {\tt PyMC3}.
At each set of model parameters the likelihood is calculated via a Gaussian
likelihood function, and multiplied by a prior (described below) to produce
the posterior probability: the probability of those model parameters given the
data.

For computational efficiency, we used {\tt PyMC3} to sample the posterior PDFs
of stellar velocities \citep{pymc3}.
This required that we rewrite the {\tt astropy} coordinate transformation code
using {\tt numpy} and {\tt Theano} \citep{numpy, theano}.
The series of rotations and translations required to convert from equatorial
to Galactocentric coordinates is described in the astropy
documentation\footnote{
    https://docs.astropy.org/en/stable/coordinates/galactocentric.html }
\citep{astropy2018}.
For each star in the \kepler\ field, we explored the posteriors of the four
parameters, \vx, \vy, \vz, and $\ln(D)$ using the {\it PyMC3} No U-Turn
Sampler (NUTS) algorithm, and the {\tt exoplanet} \python\ library
\citep{exoplanet}.
We tuned the {\it PyMC3} sampler for 1500 steps, with a target acceptance
fraction of 0.9, then ran four chains of 1000 steps for a total of 4000 steps.
This resulted in a $\hat{r}$-statistic (the ratio of intra-chain to
inter-chain variance) of around unity, indicating convergence.
Using PyMC3 made this inference procedure exceptionally fast -- taking just a
few seconds per star on a laptop.

\subsection{The prior}
\label{sec:prior}

\begin{figure}[ht!]
\caption{
\x, \y\ and \z\ positions of stars observed by \kepler, showing the
    orientation of the \kepler\ field.
The Sun's position is indicated with a Solar symbol.
The direction of the field is almost aligned with the \y-axis and almost
    perpendicular to the \x\ and \z-axes, which is why \vx\ and \vz\ can be
    tightly constrained for \kepler\ stars without RVs, but \vy\ cannot.
}
  \centering
    \includegraphics[width=.7\textwidth]{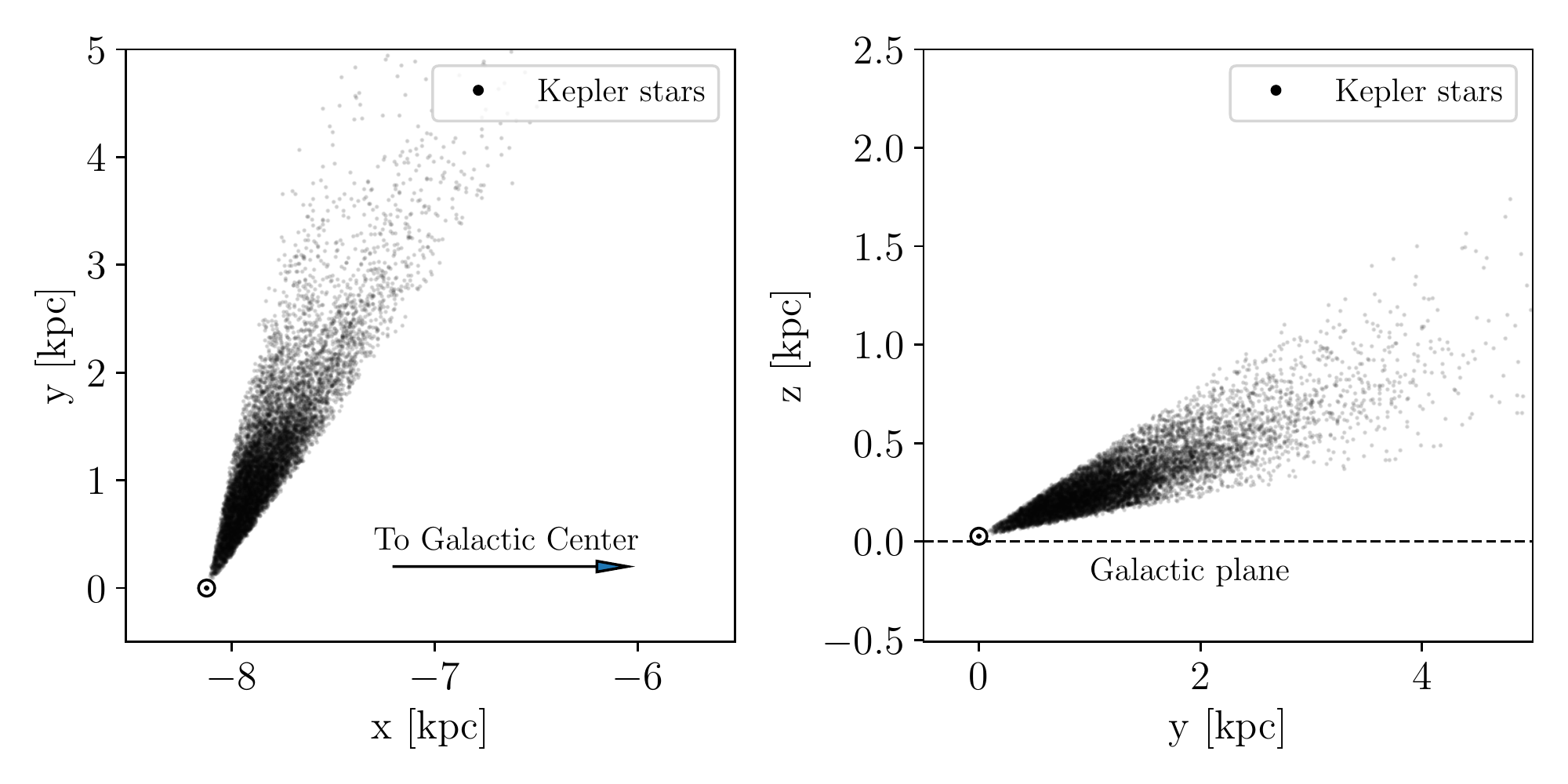}
\label{fig:kepler_field}
\end{figure}

As mentioned previously, the positioning of the \kepler\ field at low Galactic
latitude allows \vz\ to be well-constrained from proper motion measurements
alone.
This also happens to be the case for \vx, because the direction of the
\kepler\ field is almost aligned with the \y-axis of the Galactocentric
coordinate system and is almost perpendicular to both the \x\ and \z-axes (see
figure \ref{fig:kepler_field}).
For this reason, the \y-direction is similar to the radial direction for
observers near the Sun, so \vy\ will be poorly constrained for \kepler\ stars
without RV measurements.
On the other hand, \vx\ and \vz\ are almost perpendicular to the radial
direction and can be precisely inferred with proper motions alone.

\begin{figure}[ht!]
\caption{
The velocity and distance distributions for stars with RV measurements,
    used to construct a multivariate Gaussian prior over velocity and
    distance parameters for stars {\it without} RVs.
The 1- and 2-D distributions of the data are shown in black and the prior is
    indicated in red.
}
  \centering
    \includegraphics[width=.8\textwidth]{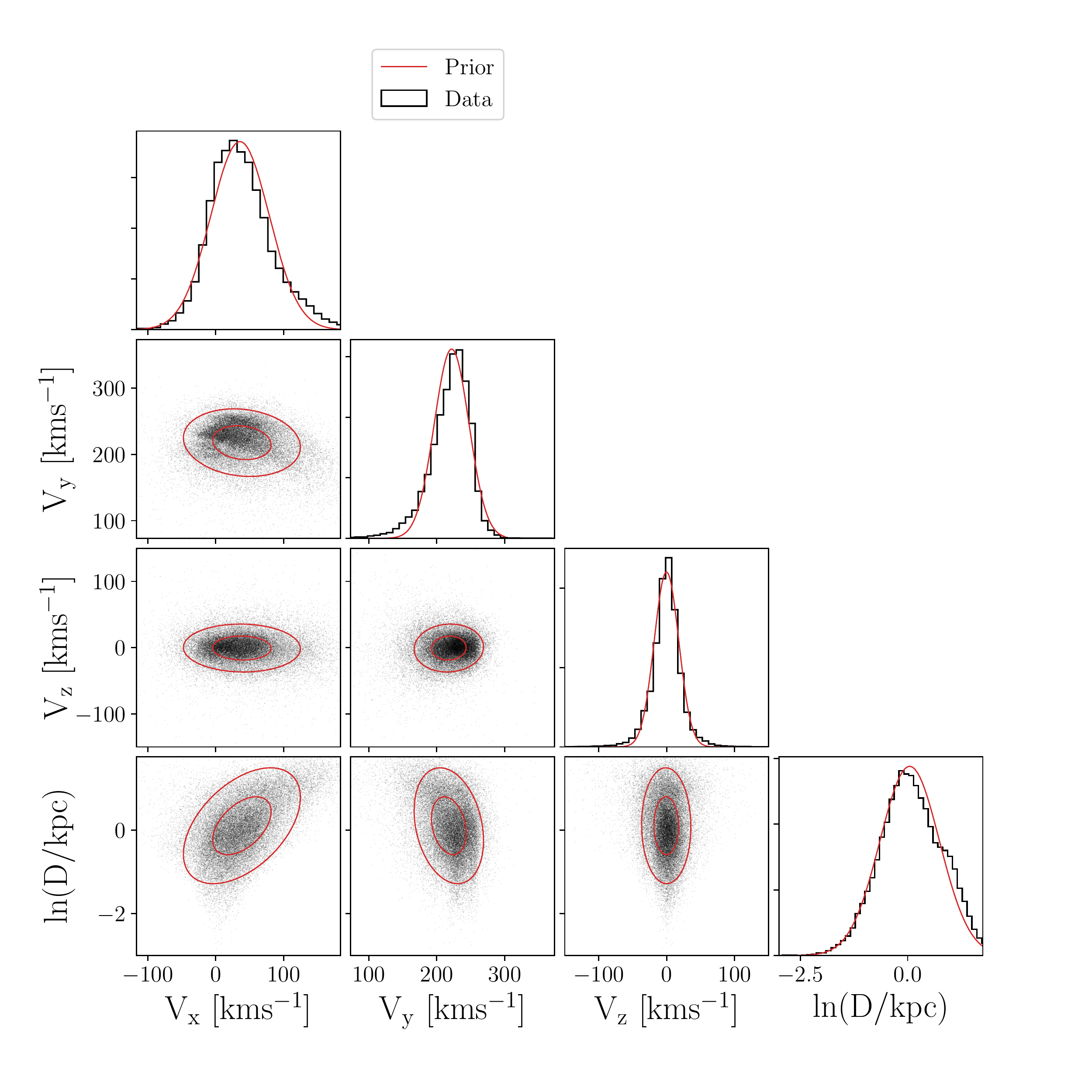}
\label{fig:prior_distributions_2D}
\end{figure}

We constructed a multivariate Gaussian prior PDF over distance and 3D velocity
using the Kepler targets {\it which have RV measurements}.
We calculated the means and covariances of the \vx, \vy, \vz\ and $\ln(D)$
distributions of stars with measured RVs and then used these means and
covariances to construct a multivariate Gaussian prior over the velocity and
distance parameters for stars {\it without} RVs.
Velocity and distance outliers greater than 3-$\sigma$ were removed before
calculating the means and covariances of the distributions.
The distance and velocity distributions of Kepler targets with RVs are
displayed in figure \ref{fig:prior_distributions_2D}.
These are the distributions we used to construct the prior.
The 1- and 2-$\sigma$ contours of the multivariate Gaussian prior is shown in
each panel in red.
This figure shows that Gaussian functions only approximately reproduce
the true velocity distributions, and do not capture the substructure.
We could have chosen a more complex prior that would fit these data better,
for example, a mixture of Gaussians, which would capture the moving groups in
the Solar neighborhood.
This may result in slightly more accurate inferred velocities.
However, since our goal is kinematic age dating, we only need to resolve the
vertical velocity component to a sufficient precision that will accurately
allow for calculations of vertical velocity dispersions, so the minor gain in
precision will not have a large affect on the end results.
In addition, since this prior is constructed using stars with RVs, which may
have a slightly different velocity distribution to stars without RVs, we opted
for the more uninformative, simple Gaussian prior.

Our goal was to infer the velocities of stars {\it without} RV measurements
using a prior calculated from stars {\it with} RV measurements.
However, stars with and without RVs are likely to be quite different
populations, determined by the Gaia, LAMOST and APOGEE selection functions.
In particular, stars without RV measurements are more likely to be fainter,
less luminous, cooler and potentially older.
Figure \ref{fig:CMD} shows the populations of stars with and without RVs on
the CMD -- stars with RVs are more likely to be upper-main-sequence and red
giant stars, and stars without RVs are more likely to be mid and lower
main-sequence dwarfs.
% Lower-mass stars are, on average, older, and have larger velocity dispersions,
% plus stars in different locations in the Galaxy have different orbital
% velocities.
For this reason, a prior based on the velocity distributions of stars {\it
with} RVs will not necessarily reflect the velocities of those without.
We could have opted to construct a prior that depends on CMD position,
however, in practice, this would require making a number of arbitrary choices,
so we instead opted for a simpler approach.
In addition, we find that the \vx\ and \vz\ velocities we infer are not
strongly influenced by the prior, as described below.
% However, given that \vx\ and \vz\ are strongly informed by proper motion
% measurements, and therefore likely to be relatively prior-insensitive, the
% prior may not significantly impact our final vertical velocities.

\begin{figure}[ht!]
\caption{
    Velocity and distance distributions of faint (blue) and bright (orange)
    stars with RVs, shown as KDEs.
    Gaussian fits to these distributions are shown as dashed lines in
    corresponding colors.
    The solid black line shows the Gaussian fit to all data (bright and faint
    combined) and is the prior we ended up using in our model.
    This figure highlights the differences between the velocities and
    distances of bright and faint stars (with RVs) in our sample.
}
  \centering
    \includegraphics[width=1\textwidth]{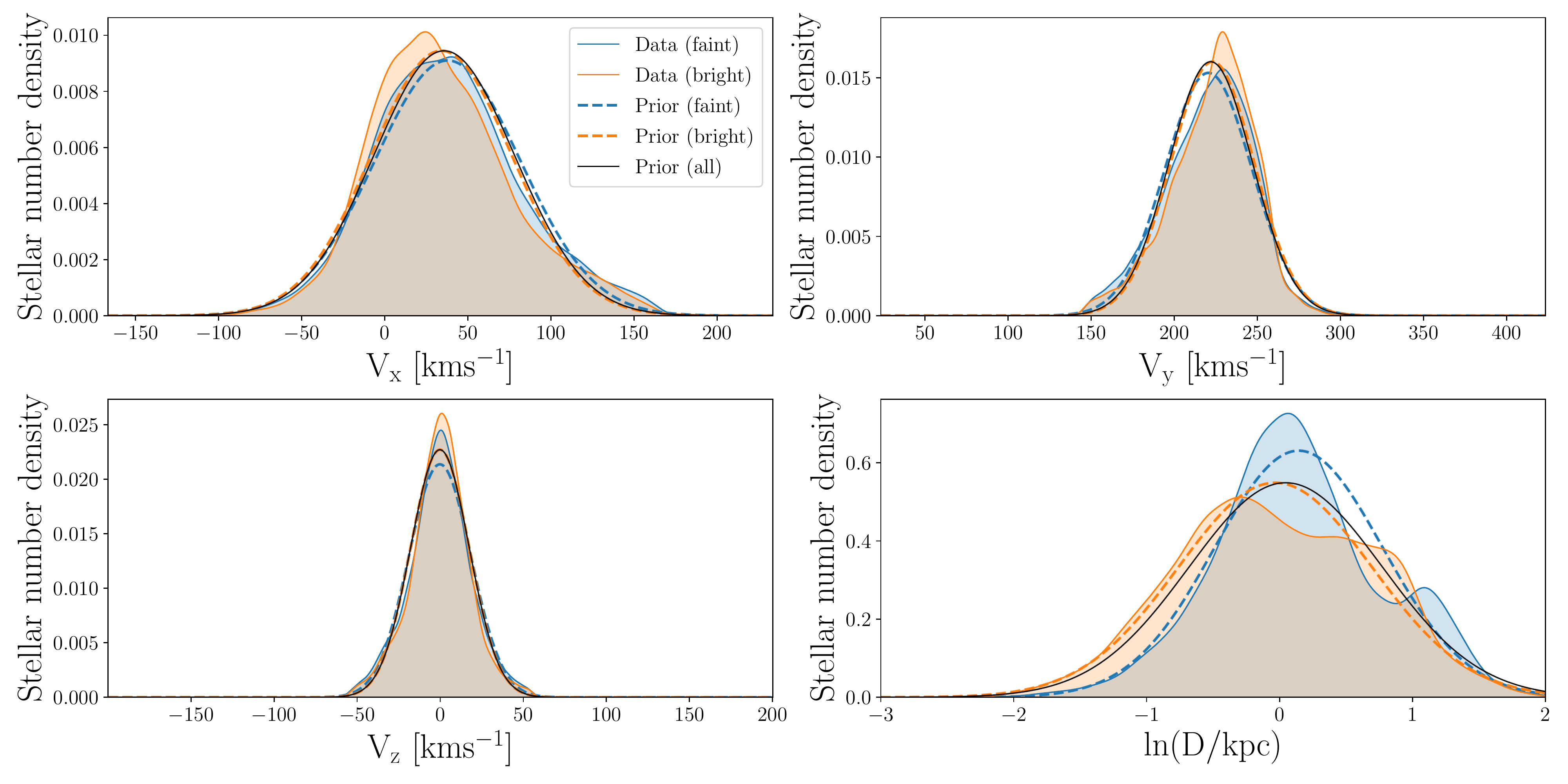}
\label{fig:prior_distributions}
\end{figure}

We tested the influence of the prior on the velocities we inferred.
One of the main features of the RV selection functions is brightness: Gaia DR2
RVs are only available for stars brighter than around 14th magnitude, and
LAMOST DR5 and APOGEE DR16 RVs for stars brighter than around 16th magnitude.
For this reason, we tested priors based on stellar populations with different
apparent magnitudes.
Three priors were tested: one calculated from the velocity distributions of
the brightest half of the RV sample (\gaia\ $G$-band apparent magnitude $<$
13), one from the faintest half ($G$ $>$ 13), and one from {\it all} stars
with RVs.
Figure \ref{fig:prior_distributions} shows the distributions of the faint
(blue) and bright (orange) halves of the RV sample as kernel density estimates
(KDEs).
The distributions are different because bright stars are typically more
massive, younger, more evolved, and/or closer to the Sun on average than faint
stars.
As a result, these stars occupy slightly different Galactic orbits.
The multivariate Gaussian, fit to these distributions, which was used as a
prior PDF, is shown as single-dimension projections in figure
\ref{fig:prior_distributions}.
The Gaussian fit to the bright and faint star distributions are shown as
dashed orange and blue lines, respectively.
The Gaussian fit to {\it all} the data, both bright and faint, is shown as a
black solid line.
The means of the faint and bright distributions differ by 6 \kms, 5 \kms, 1
\kms\ and 0.21 kpc, for \vx\, \vy, \vz\ and $\ln(D)$, respectively.
The \vx, \vy, and distance distributions of the bright stars are slightly
non-Gaussian -- more so than the faint stars.
This highlights the inadequacy of using a Gaussian distribution as the prior
-- a Gaussian is only an approximation of the underlying distribution of stars
in our sample.
As a result of this approximation, inferred velocities that are strongly
prior-dependent -- (\ie\ especially those in the \y-direction) may inherit
some inaccuracies from the Gaussian prior, which is not a perfect
representation of the underlying data.
However, given that the populations of stars with and without RV measurements
are different, it may be inappropriate to use a more complex, more
informative prior anyway.

\begin{figure}[ht!]
\caption{
The distributions of velocity and distance parameters, inferred using three
    different priors.
The orange line is a KDE that represents the distribution of parameters
    inferred with a Gaussian prior, estimated from the bright half of the RV
    sample ($G < $ 13).
The blue dashed line shows the results from a prior estimated from the faint
    half of the RV sample ($G > 13$)).
The black line shows the results from a prior calculated from all stars with
    RV measurements and is the prior we adopt in our final analysis.
    This figure shows that all parameters except \vy\ are relatively
    insensitive to the three priors that were tested.
    }
  \centering
    \includegraphics[width=1\textwidth]{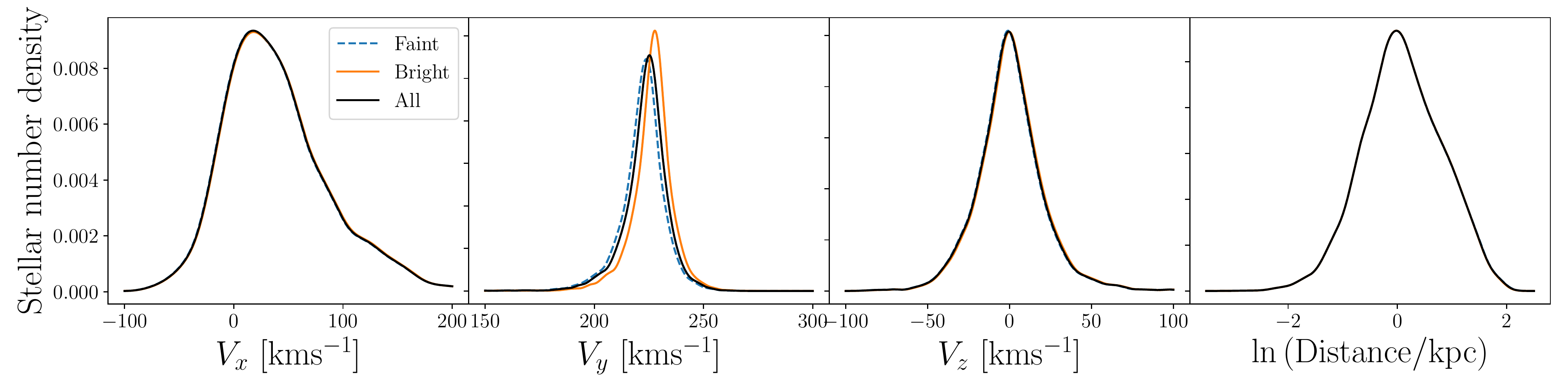}
\label{fig:prior_comparison}
\end{figure}

We inferred the velocities of 1000 stars chosen at random from the
RV Kepler sample using each of these three priors and compared the
inferred velocity distributions.
If the inferred velocities were highly prior-dependent, the resulting
distributions, obtained from different priors, would look very different.
The results of this test are shown in figure \ref{fig:prior_comparison}.
From left to right, the three panels show the distributions of inferred \vx,
\vy, \vz, and log-distance.
The blue dashed line shows a KDE representing the distributions of velocities
inferred using the prior calculated from the faint half of the RV sample.
Similarly, the solid orange line shows the distribution of inferred velocities
using the prior calculated from the bright half of the RV sample, and the
solid black line shows the results of the prior calculated from {\it all}
stars with measured RVs.
In all but the \vy\ panel (second from the left), the blue, orange,
and black lines lie on top of each other, indicating that different priors do
not significantly influence the resulting velocities.

The median values of the \vy\ distributions resulting from the faint and
bright priors differ by around 4 \kms.
This is similar to the difference in means of the faint and bright populations
(5 \kms, as quoted above).
The inferred \vx\ and \vz\ distributions differ by 2 \kms\ and 0.3 \kms,
respectively.
Regardless of the prior choice, the \vx\ and \vz\ distributions are similar
because velocities in the \x\ and \z-directions are not strongly prior
dependent: they are tightly constrained with proper motion measurements alone.
However, the distribution of inferred \vy\ velocities {\it does} depend on the
prior.
This is because the \y-direction is close to the radial direction for \kepler\
stars (see figure \ref{fig:kepler_field}), and \vy\ cannot be tightly
constrained without an RV measurement.
The distributions of stellar distances are almost identical, irrespective of
the prior.
This is because distance is very tightly constrained by Gaia parallax and is
relatively insensitive to the prior.
% It is therefore highly dependent on the prior.

% Fainter stars have smaller y-velocities than brighter stars because they are,
% on average, further from the Sun

Although this test was performed on stars with RV measurements, which are
brighter overall than the sample of stars without RVs (\eg\ figure
\ref{fig:rv_histogram}), figure \ref{fig:prior_comparison} nevertheless
indicates that \vx\ and \vz\ are not strongly prior-dependent.
Since this work is chiefly motivated by kinematic age-dating, which mostly
requires vertical velocities (\vz), we are satisfied with these results.
% The difference in the dispersions of \vz\ velocities, calculated with the
% three different priors tested above was smaller than 0.5 \kms.
We conclude that the \vx\ and \vz\ velocities we infer are relatively
insensitive to prior choice, and we adopt a prior calculated from the
distributions of all stars with RV measurements (black Gaussians in figure
\ref{fig:prior_distributions}).
The \vy\ velocities are more strongly prior dependent and should be used with
caution.

A better prior for the velocity distribution would be a phase-space
distribution function that takes into account asymmetric drift and other
nontrivial aspects of the local velocity distribution.
However, priors of this form would require a model for the phase-space
distribution function, $f(x,y,z,v_x,v_y,v_z)$, which would require a
hierarchical model that takes into account covariances between the spatial and
kinematic properties of stars \citep[\eg][]{trick2016, hagen2019,
anguiano2020}.
We consider this to be outside of the scope of this work, and instead argue
that our assumed isotropic velocity prior is a conservative choice because it
does not impose any covariant structure between velocity components or
dependencies on Galactic position.
However, in general, to extend this Kepler-field-specific analysis to other
populations of stars in different parts of the Galaxy, alternative priors will
have to be constructed.
We leave this for a future exercise.

\begin{figure}[ht!]
\caption{
The posterior PDF over parameters \vx, \vy, \vz\ and $\ln$(distance) for a
    Kepler target chosen at random: KIC \kicstar.
    Blue lines show the location of the star's `true' parameters, based on a
    calculation that includes the RV measurement.
This figure shows that the velocity parameters are correlated and the star's
    posterior is elongated in \vy.
    }
  \centering
    \includegraphics[width=.7\textwidth]{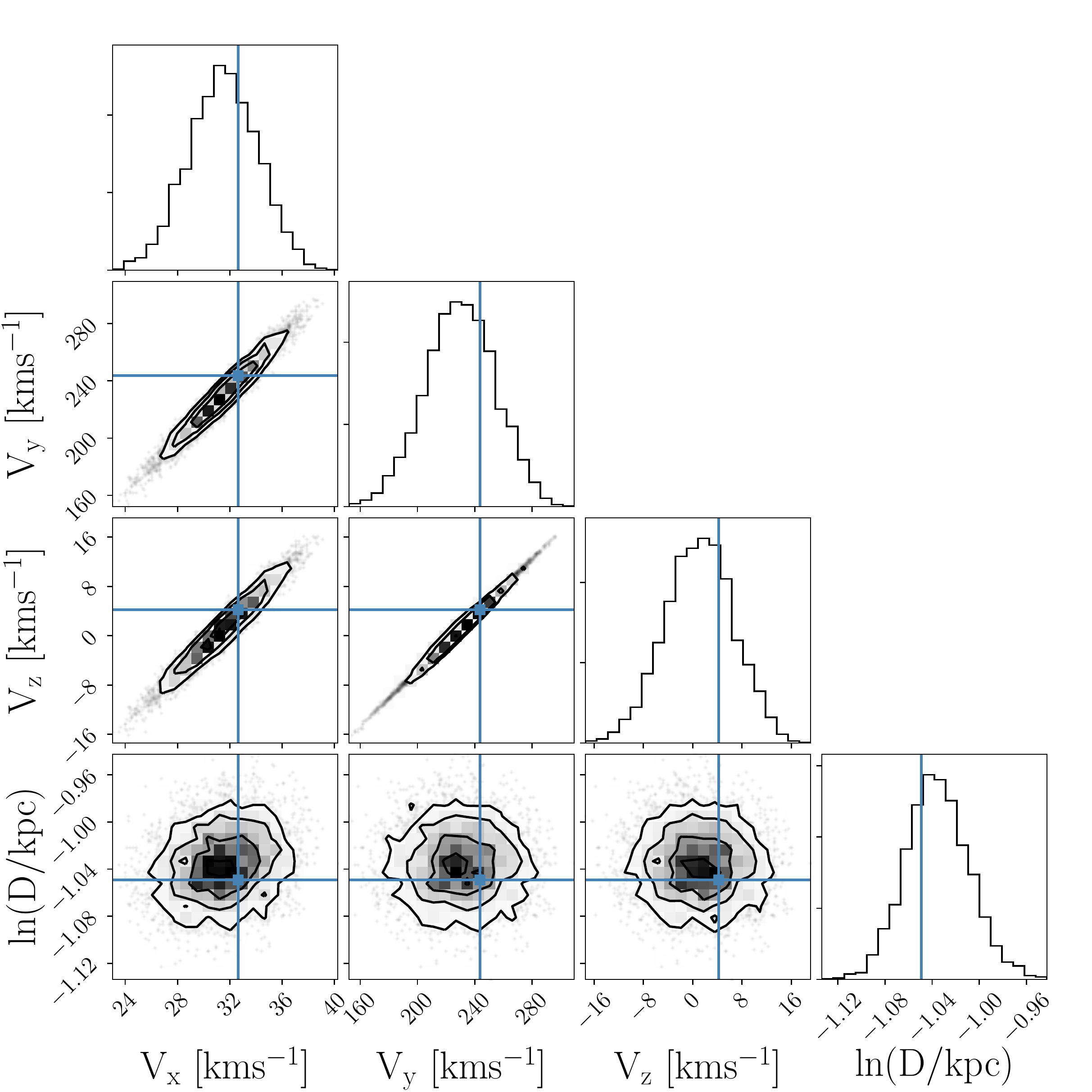}
\label{fig:posterior}
\end{figure}

Figure \ref{fig:posterior} shows the posterior PDF over velocity and distance
parameters for a randomly selected Kepler target with an RV measurement, KIC
\kicstar.
The blue lines in each panel indicate star's velocities, directly calculated
using the RV measurement, and the distributions indicate the probability
density of the parameters inferred {\it without} the RV measurement.
The velocity parameters are correlated because the lack of RV introduces a
slight degeneracy: the star's proper motion can be equally well described with
a range of different velocities.
The star's posterior PDF is particularly elongated in \vy, which is the
velocity most similar to RV.
The star's distance is not tightly correlated with its velocity parameters
because it is precisely determined by parallax.

\section{Results and Discussion}
\label{sec:results}

\subsection{Inferred velocities}

\begin{figure}[ht!]
\caption{
The distribution of inferred stellar velocities and distances.
    Figure \ref{fig:results} shows the inferred velocities of 5000 randomly
selected Kepler stars.
The 2D distributions of inferred stellar velocities are plotted in the
lower-left panels, with black contours indicating the stellar number density.
The red contours in the lower-left panels show the marginal projections of
    the Gaussian prior distribution in 2D.
The upper-right panels in the figure, lying on the plot's diagonal, show the
    1D distributions (histograms) of stellar velocities.
The black histogram shows the distribution of inferred velocities, the blue
histogram shows the distribution of velocities for stars with RVs,
and the red lines show the 1D marginal Gaussian prior distributions.
This figure shows that the distributions of inferred velocities are
    broadly similar to the distributions of velocities directly calculated
    using RVs, and the prior constructed from them.
}
  \centering
    \includegraphics[width=.7\textwidth]{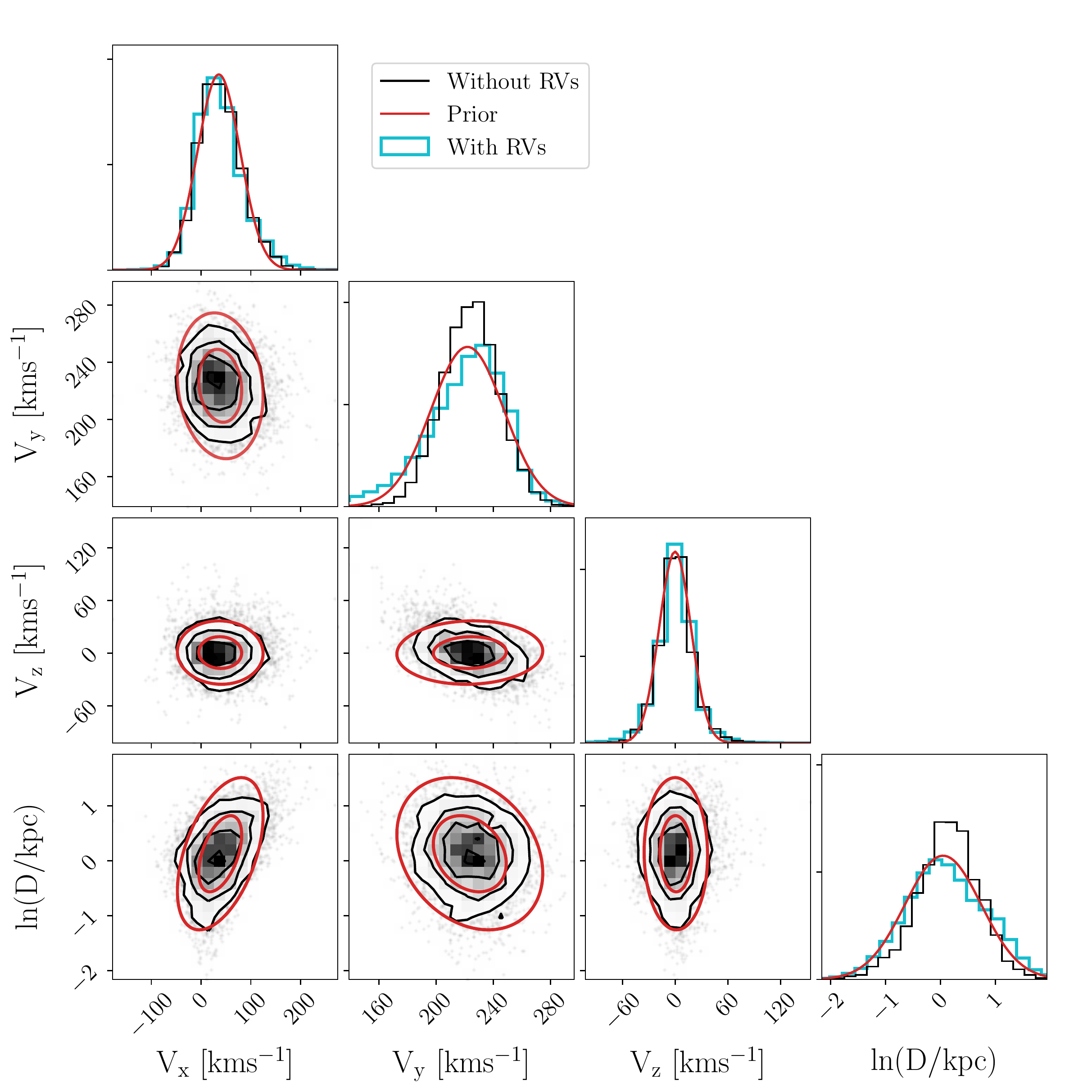}
\label{fig:results}
\end{figure}

In this section we assess the quality of the 3D stellar velocities we infer.
Figure \ref{fig:results} shows the distribution of stellar velocities inferred
for 5000 randomly selected Kepler stars.
The 2D distributions of inferred stellar velocities are plotted in the
lower-left panels, with black contours indicating the stellar number density.
The red contours in these panels show the marginal projections of the
Gaussian prior in 2D.
The diagonal panels show the 1D distributions (histograms) of stellar
velocities.
The black histogram shows the distribution of inferred velocities, the cyan
histogram shows the distribution of velocities calculated for stars with RVs
(on which the prior was based), and the red lines show the 1D prior
distributions.
The prior distribution was calculated using the velocities of stars with RVs.
If the velocity distributions of stars were Gaussian, the 1D red Gaussians
would look like the cyan histograms.
In other words, the differences between the red lines and cyan histograms
is caused by the non-Gaussianity of the velocity distributions.
This figure is intended to highlight the differences and similarities
between the inferred stellar velocities and the prior distributions.
In each panel, the distribution of inferred velocities is fairly similar to
the distribution of directly-measured velocities; the velocities of stars
calculated with and without RVs are broadly similar.
The inferred \vx\ and \vz\ velocities, and distances are relatively
precise, with median uncertainties of \vxprecision \kms, \vzprecision\ \kms,
and \dprecision\ kpc, respectively.
The inferred \vy\ velocities have a median uncertainty of \vyprecision\ \kms.

% Among stars with measured RVs, \vy\ and \vz\ are slightly positively
% correlated, \ie\ stars with larger \vy\ tend to have larger \vz.
There is a slight negative correlation between inferred \vy\ and \vz\
velocities, which is visible in the central panel of figure \ref{fig:results}.
This negative correlation is not seen in the prior, nor is it apparent in the
posteriors of individual stars: figure \ref{fig:posterior} shows that the \vy\
and \vz\ parameters are {\it positively} correlated for KIC \kicstar.
This negative correlation may be due to the specific orientation of the Kepler
field, which creates a slight degeneracy between \vy\ and \vz.
% , and could
% result in a negative correlation in the population of stars that is not
% apparent in the posteriors of individual stars.
% To an observer looking at Kepler field, a star with either a positive
% \vz\ or a {\it negative} \vy\ would appear to move in the direction of
% positive \vz, when projected onto the sky.
% In this sense, \vy\ and \vz\ are negatively correlated.
% However, the observed proper motions of a star without a measured RV could be
% equally well described by either increasing both \vy\ and \vz, or decreasing
% both \vy\ and \vz.
% For this reason, the star's posterior PDF over \vy\ and \vz\ will be
% positively correlated.
% This could result in a paradoxical result whereby the posteriors over vy and
% vz are positively correlated for individual stars, however the vy and vz
% velocities of the population are negatively correlated.
If this explanation is correct, this phenomenon highlights how apparent
correlations in stellar velocities could result from {\it missing}
information.
This point is interesting, but the effect is relatively small and we do not
expect that this observed correlation between \vz\ and \vy\ will significantly
affect kinematic age studies.

% could have {\it
% either} a positive \vz, or a postitive \vy. with a positive
% \vz\
% This difference is caused by the way the inaccuracies in inferred \vy\
% velocities manifest in the \vz\ velocity distribution.
% Although correlation patterns appear between the inferred velocities as a
% result of marginalizing over RV, the inferred velocities are still consistent
% with their `true' velocities.
% The proper motions of stars with a given \vy\ and \vz\ could be equally
% well-described with a slightly larger \vy\ and a smaller \vz\ or vice versa.

\begin{figure}[ht!]
\caption{Velocities calculated with full 6D information compared with
    velocities inferred without RVs, for 5000 Kepler targets with RV
    measurements.}
  \centering
    \includegraphics[width=1\textwidth]{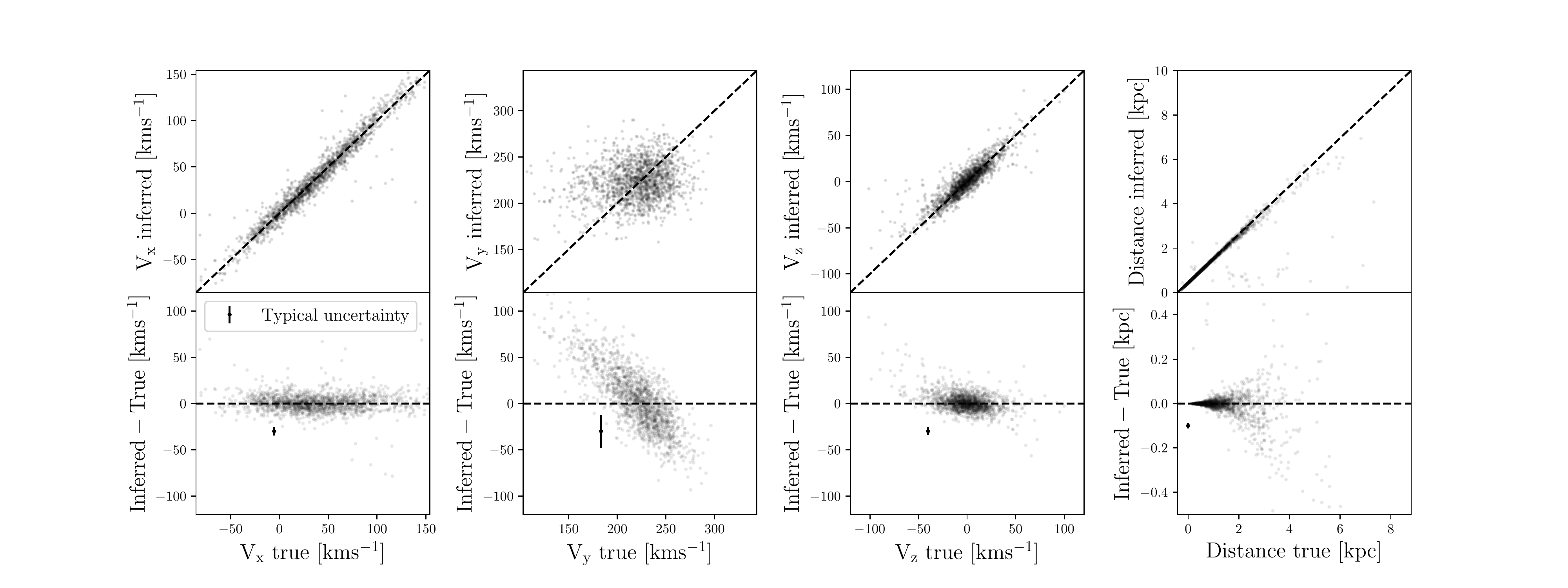}
\label{fig:residuals}
\end{figure}

To further validate our method, we compare inferred velocities with
directly-calculated velocities for stars in our sample with measured RVs.
Figure \ref{fig:residuals} shows the \vx, \vy\ and \vz\ velocities, and
distances we inferred, compared with those calculated from measured RVs, for
5000 Kepler stars chosen at random.
The three velocity components, \vx, \vy\ and \vz\ were recovered with
differing levels of precision: \vx\ and \vz\ are inferred more precisely than
\vy.
This is because of the position of the \kepler\ field, shown in figure
\ref{fig:kepler_field}.
The velocities of low-\vy\ stars are overestimated and the velocities of
high-\vy\ stars are underestimated.
This is because there is little information to constrain the \vy\ velocities
and the prior pulls the \vy\ velocities toward the center of the distribution.
The \vx, \vy\ and \vz\ velocities of stars are correlated, which means that
stars with an inaccurate \vy\ also have slightly accurate \vz\ and \vx.
Despite slight systematic inaccuracies visible in the residual (bottom)
panels of figure \ref{fig:residuals}, around 68\% of the inferred velocities
are within 1$\sigma$ of their true velocities; the inferred velocities are
consistent with the true velocities.

We provide a table of the directly-calculated, and indirectly-inferred 3D
velocities of stars observed by Kepler, in addition to their positional and
velocity information from Gaia EDR3, LAMOST DR5 and APOGEE DR16.
A description of each column included in that table is provided in table
\ref{tab:columns}.
% A sample of this table is displayed here, and the full machine-readable table
% is available online.

\begin{figure}[ht!]
\caption{
Residual velocity distributions calculated by marginalizing over missing RV
    measurements (green) or by assuming the RV can be set to zero (purple) for
    5000 stars in our sample.
Each distribution shows the residuals between the inferred/calculated
    velocities, and the velocities calculated using full 6D information:
    parallaxes, positions, proper motions, and RVs.
The vertical lines show the medians of each distribution.
This figure demonstrates that calculating stellar velocities using proper
    motions and distances alone will significantly bias the results.
Marginalizing over missing RV measurements ensures the resulting velocities
    will be unbiased.
}
  \centering
    \includegraphics[width=1\textwidth]{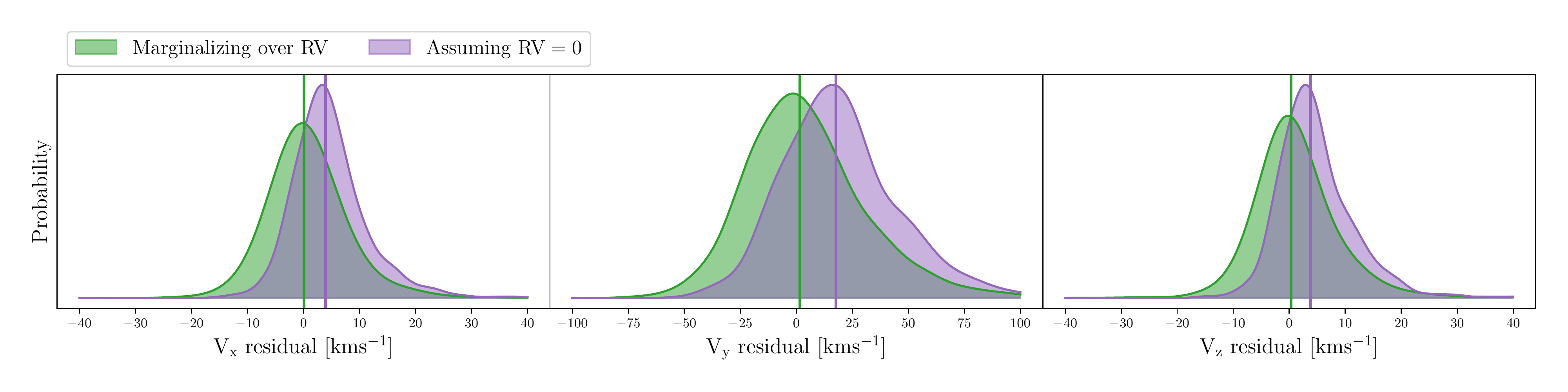}
\label{fig:inferred_vs_calc}
\end{figure}

As briefly mentioned in the introduction, we have gone to the trouble of
inferring velocities in this paper because calculating velocities from proper
motions and distances alone, \ie\ assuming the RV is zero, will bias the
resulting velocities.
To demonstrate this, figure \ref{fig:inferred_vs_calc} shows the velocity
residuals for stars in our sample where we infer the velocities by
marginalizing over RV and where we ignore RV, \ie\ set it to zero.
In each panel, the green distributions show the residuals between the true
velocities and our inferred velocities, and the purple distributions show the
residuals between the true velocities and velocities calculated by setting the
RV to zero.
The median of each distribution is shown as a vertical line.
This figure shows that ignoring the RV dimension and assuming it can be set to
zero significantly biases the resulting velocities.
The correct way to de-bias velocity calculations is to marginalize over
missing RV measurements.

Our intention in providing a table of stellar velocities, and outlining our
method for inferring velocities without RV measurements, is to facilitate
kinematic age-dating studies.
The vertical velocities we provide could be used as an age proxy, or to
calculate kinematic ages via an age-velocity dispersion relation
\citep[\eg][]{yu2018, mackereth2019, sharma2021}.
This kind of approach has been used many times to examine the kinematic ages
of cool stars to explore stellar and planetary evolution
\citep[\eg][]{newton2016, kiman2019, hamer2019, angus2020, lu2021}.
For example, \citet{kiman2019} used the vertical velocity dispersions of stars
as an age proxy to explore the evolution of H$\alpha$ equivalent width (a
magnetic activity indicator), in M dwarfs.
\citet{hamer2019} compared the vertical velocity dispersions of stars with and
without hot Jupiters to show that hot Jupiter hosts tend to be younger, on
average, than stars without detected hot Jupiters.

% \begin{table}[h!]
%   \begin{center}
%       \caption{
% The velocities and kinematic information for the \nstars\ Kepler targets in
%       our sample.
%       The full list of columns is provided in table \ref{tab:columns} and the
%       full version is published in its entirey online.
%       }
% \label{tab:data}
% \begin{tabular}{cccccccccc}
%     KIC ID & DR3 source ID & $\alpha$ & $\delta$ & $\bar{\omega}$
%     & Distance & $\mu_{\alpha}$ & $\mu_\delta$ & RV (Gaia) & ... \\
%     & & $(^\circ)$ & $(^\circ)$ & mas & kpc & mas/yr & mas/yr & km/s & \\
%     {\tt kic$\_$id} & {\tt source$\_$id} & {\tt ra} & {\tt dec} &
%     {\tt parallax} & {\tt r$\_$est} & {\tt pmra} & {\tt pmdec} \\
% \hline
% ... & ... & ... & ... & ... & ... \\
% \end{tabular}
% \end{center}
% \tablecomments{This table is published in its entirety in a machine-readable format.
%       A portion is shown here for guidance regarding its form and content.}
% \ref{tab:data}
% \end{table}

\begin{table}[h!]
  \begin{center}
      \caption{
The list of columns in the data table published online, which provides
      the velocities and kinematic information for stars in the Kepler field.
      }
\label{tab:columns}
\begin{tabular}{cc}
    Column name & Description \\
\hline
    {\tt KIC} & The Kepler Input Catalog ID number of the target. \\
    {\tt Gaia} & The Gaia DR3 source ID of the target. \\
    {\tt RAdeg, e$\_$RAdeg} & Gaia EDR3 right ascenscion ($^\circ$). \\
    {\tt DEdeg, e$\_$DEdeg} & Gaia EDR3 declination in degrees ($^\circ$). \\
    {\tt plx, e$\_$plx} & Gaia EDR3 parallax (mas). \\
    {\tt Dist, b$\_$Dist, B$\_$Dist} & Distance, lower and upper bounds
    (parsec), provided by \citet{bailer-jones2021}. \\
    {\tt pmRA, e$\_$pmRA} & Gaia EDR3 proper motion in right ascension (mas/yr). \\
    {\tt pmDE, e$\_$pmDE} & Gaia EDR3 proper motion in declination (mas/yr). \\
    {\tt RV-gaiaDR2, e$\_$RV-gaiaDR2} & Gaia DR2 radial velocity (km/s) \\
    {\tt RV-apo16, e$\_$RV-apo16} & APOGEE DR16 radial velocity (km/s) \\
    {\tt RV-lamDR5, e$\_$RV-lamDR5} & LAMOST DR5 radial velocity (km/s) \\
    {\tt vx-calc} & The \vx\ velocity calculated using RV (km/s). \\
    {\tt vx-inferred, e$\_$vx-inferred} & Median and std.
    dev, of \vx\ velocity samples, inferred without RV (km/s). \\
    {\tt vy-calc} & The \vy\ velocity calculated using RV (km/s). \\
    {\tt vy-inferred, e$\_$vy-inferred} & Median and std.
    dev. of \vy\ velocity samples, inferred without RV (km/s). \\
    {\tt vz-calc} & The \vz\ velocity calculated using RV (km/s). \\
    {\tt vz-inferred, e$\_$vz-inferred} & Median and std.
    devi. of \vz\ velocity samples, inferred without RV (km/s). \\
    {\tt vxvy-covar} & The covariance between \vx\ and \vy\ samples. \\
    {\tt vxvz-covar} & The covariance between \vx\ and \vz\ samples. \\
    {\tt vxlnd-covar} & The covariance between \vx\ and $\ln$(distance)
    samples. \\
    {\tt vyvz-covar} & The covariance between \vy\ and \vz\ samples. \\
    {\tt vylnd-covar} & The covariance between \vy\ and $\ln$(distance)
    samples. \\
    {\tt vzlnd-covar} & The covariance between \vz\ and $\ln$(distance)
    samples. \\
\end{tabular}
\end{center}
\end{table}

\section{Conclusion}

This paper describes a method for inferring the 3D velocities of stars by
marginalizing over missing radial velocity measurements.
We focused on stars in the Kepler field because of its potential for studying
stellar evolution via kinematic age-dating as well as its advantageous
orientation.
Located at low Galactic latitude, the Kepler field is almost aligned with the
$y$-axis of the Galactocentric coordinate system.
This means that 2D Gaia proper motion measurements alone are sufficient to
tightly constrain the \vx\ and \vz\ velocities of Kepler stars.
Without RV measurements, the \vy\ velocities  of Kepler stars are
poorly constrained.
However, given that many age-velocity dispersion relations (AVR) are
calibrated in {\it vertical} velocity, \vz\ is the main parameter of interest
for kinematic age-dating and it is precisely constrained by our method: \vz\
is inferred with a median precision of \vzprecision\ kms$^{-1}$.

We compiled kinematic data for Kepler targets from the, Gaia EDR3, LAMOST DR5
and APOGEE DR16 catalogs.
Gaia EDR3 provided parallaxes, positions and proper motions for the stars in
our sample.
Altogether, Gaia DR2, LAMOST DR5, and APOGEE DR16 provided RVs for \nrv\
Kepler targets.
% Of the three spectroscopic surveys, APOGEE has the highest resolution,
% followed by Gaia, then LAMOST, so we adopted RVs in that priority-order
% where stars had multiple RV measurements available.

We calculated \vx, \vy, and \vz\ for the \nrv\ stars in our sample with RVs
using {\tt astropy}.
For the remaining stars, we {\it inferred} \vx, \vy, \vz, and distance while
marginalizing over RV.
Our prior was a 4D Gaussian in \vx, \vy, \vz\ and $\ln$(distance), which was
based on the distribution of stars in our sample {\it with} RVs.
Since the populations of stars with and without RVs in the Kepler field are
somewhat different -- stars {\it with} RVs are generally brighter than stars
without -- we tested the sensitivity of our results to the prior.
We split the subsample of stars with measured RVs into two further subgroups:
stars brighter and stars fainter than 13th magnitude in Gaia $G$-band (13th
being the median magnitude of the Kepler stars with RVs).
Priors were constructed from the faint and bright halves of the sample and
used to infer the velocities of 1000 stars randomly selected from the total RV
sample.
Upon examination, we found the final inferred velocities were similar,
irrespective of the prior.
As expected, \vx\ and \vz\ depend very little on the prior but \vy\ has a
stronger prior-dependence because it is difficult to constrain without an RV
for Kepler stars.
A caveat of our inferred velocities is therefore that the \vy\ velocities may
not be accurate for faint stars in the Kepler field.
% Overall, we found that the distribution of inferred velocities for Kepler
% targets is similar to the velocity distribution of stars with measured RVs.
The median precision of inferred \vx, \vy, and \vz\ velocities is
\vxprecision, \vyprecision, and \vzprecision\ kms$^{-1}$ respectively.
We provide a table of parameters \vx, \vy, \vz, and $\ln$(distance), with
uncertainties and covariances, for a total of \nstars\ Kepler targets.
This table also contains the positional and velocity information from Gaia
DR2, Gaia EDR3, LAMOST DR5, and APOGEE DR16 used in this project.

\section*{Acknowledgements}

% This work was partly developed at the 2019 KITP conference `Better stars,
% better planets'.
% Parts of this project are based on ideas explored at the Gaia sprints at the
% Flatiron Institute in New York City, 2016 and MPIA, Heidelberg, 2017.
The authors would like to thank the anonymous referee whose helpful
suggestions significantly improved this manuscript.

This work made use of the gaia-kepler.fun crossmatch database created by Megan
Bedell.

Some of the data presented in this paper were obtained from the Mikulski
Archive for Space Telescopes (MAST).
STScI is operated by the Association of Universities for Research in
Astronomy, Inc., under NASA contract NAS5-26555.
Support for MAST for non-HST data is provided by the NASA Office of Space
Science via grant NNX09AF08G and by other grants and contracts.
This paper includes data collected by the Kepler mission. Funding for the
\Kepler\ mission is provided by the NASA Science Mission directorate.

This work has made use of data from the European Space Agency (ESA) mission
{\it Gaia} (\url{https://www.cosmos.esa.int/gaia}), processed by the {\it
Gaia} Data Processing and Analysis Consortium (DPAC,
\url{https://www.cosmos.esa.int/web/gaia/dpac/consortium}).
Funding for the DPAC has been provided by national institutions, in particular
the institutions participating in the {\it Gaia} Multilateral Agreement.

RA acknowledges support from Astrophysics Data Analysis Program award ADAP
\#80NSSC21K0636.

JCZ is supported by an NSF Astronomy and Astrophysics Postdoctoral Fellowship
under award AST-2001869.

\software{Astropy \citep{astropy2013, astropy2018}; Matplotlib
\citep{matplotlib}; Seaborn, \citep{seaborn}; Numpy, \citep{numpy}; Theano,
\citep{theano}; PyMC3, \citep{pymc3}; Exoplanet, \citep{exoplanet}}

\bibliography{refs}{}
\bibliographystyle{aasjournal}

\end{document}